\documentclass[%
 aip,
 amsmath,amssymb,
 reprint,%
]{revtex4-1}
\usepackage{comment}
\usepackage{graphicx}% Include figure files
\usepackage{dcolumn}% Align table columns on decimal point
\usepackage{bm}% bold math
\usepackage[justification=raggedright,singlelinecheck=false]{caption}
\usepackage[utf8]{inputenc}
\usepackage[T1]{fontenc}
\usepackage{mathptmx}
\usepackage{etoolbox}
\usepackage{xcolor}
\usepackage{soul}
\usepackage{amsmath,amssymb,amsfonts}
\usepackage{algorithmic}
\usepackage{graphicx}
\usepackage{helvet}      % Helvetica (Arial-like)
\usepackage{textcomp}
\usepackage[normalem]{ulem}
\usepackage{tikz}
\usepackage{caption}
\usepackage{subcaption}
\usepackage[utf8]{inputenc}
\usetikzlibrary{arrows.meta,calc,positioning}

\makeatletter
\def\@email#1#2{%
 \endgroup
 \patchcmd{\titleblock@produce}
  {\frontmatter@RRAPformat}
  {\frontmatter@RRAPformat{\produce@RRAP{*#1\href{mailto:#2}{#2}}}\frontmatter@RRAPformat}
  {}{}
}%
\makeatother

\begin{document}

\preprint{AIP/123-QED}

\title{Hot LO Phonon-Induced RF Nonlinearity in GaN High-Electron-Mobility Transistors}

\author{Ankan Ghosh Dastider}
\affiliation{%
Department of Electrical and Computer Engineering, University of Illinois Urbana-Champaign, Urbana, IL, USA
}%
\affiliation{%
Holonyak Micro and Nanotechnology Laboratory, University of Illinois Urbana-Champaign, Urbana, IL, USA
}%

\author{Matt Grupen}
\affiliation{ 
Air Force Research Laboratory Sensors Directorate, 2241 Avionics Cir., Wright-Patterson AFB, OH, USA
}%

\author{Nicholas C. Miller}
\affiliation{%
Department of Electrical and Computer Engineering, Michigan State University, East Lansing, MI, USA
}%

\author{Shaloo Rakheja\textsuperscript{*,}}
  \email{rakheja@illinois.edu}
\affiliation{%
Department of Electrical and Computer Engineering, University of Illinois Urbana-Champaign, Urbana, IL, USA
}%
\affiliation{%
Holonyak Micro and Nanotechnology Laboratory, University of Illinois Urbana-Champaign, Urbana, IL, USA
}%

\date{\today}

\begin{abstract}
Hot longitudinal optical (LO) phonons in GaN have recently been identified as a major factor degrading the DC performance of GaN high-electron-mobility transistors (HEMTs) by 30--60\%, despite their ultrafast decay. However, their impact on large-signal RF performance, particularly RF linearity, remains poorly understood. Using full-band transport simulations of a fabricated GaN HEMT, we show that even ultrafast LO phonons with a lifetime of 30\,fs degrade the output 1-dB compression point and the third-order output intercept power by $\sim$3~dB compared to the case without LO phonon heating. Furthermore, our analysis reveals that improvements in transconductance ($g_\textrm{m}$) flatness do not necessarily translate into improved RF linearity because multiple nonlinear mechanisms contribute to the transistor response, and their combined effect cannot be captured by $g_\mathrm{m}$ flatness alone. This work clarifies a persistent ambiguity in the literature regarding using $g_\mathrm{m}$ flatness as a proxy for RF linearity and establishes \textcolor{black}{intrinsic} phonon-induced limits on the RF performance of GaN HEMTs.
\end{abstract}

\maketitle

\section{Introduction}
\label{sec:introduction}
GaN-based high-electron-mobility transistors (HEMTs) have become essential to meet the growing need for high-power microwave applications.~\cite{teoEmergingGaN2021, palmourWideBandgapRF2001}
%~\cite{b1,b2}. 
With the continued evolution of wireless communication systems, stringent demands are imposed on the output power and linearity performance of RF power amplifiers.~\cite{nagyLinearityGaNHEMT2003,jenkinsLinearityAlGaN2001} Extensive research on GaN HEMT RF nonlinearity has produced competing ideas about its fundamental origins. For example, it has been suggested that interface roughness \cite{liMonteCarloAlGaN2000, juangTransportAlGaNGaN2003} and longitudinal optical (LO) phonon scattering \cite{bajajTransportGaNHEMT2015} may cause gate bias variations in the transconductance $g_\mathrm{m}$, leading to RF nonlinearity. Self-heating at high currents has also been considered.~\cite{chenBellShapeGM2016} Finally, $g_\mathrm{m}$ degradation by increased dynamic source access resistance at high drain currents has also been investigated as a potential cause of RF nonlinearity.~\cite{palaciosDynamicAccessResistance2005}
%{\color{green} It is interesting to note that a consistent throughline in studies like those cited above is the focus on $g_m$ as the indicator of RF nonlinearity.}

Among the various mechanisms mentioned above, the generation and accumulation of hot LO phonons~\cite{ridleyQuantumProcesses2013,ridleyHotPhononVelocity2004,barmanNonequilibriumPhonons2004,srivastavaHotPhononOrigin2008,matulionisHotPhononTemperature2003} and associated electron scattering needs special attention %This is because unlike other mechanisms linked to the RF performance, the 
because unlike other limiting factors, LO phonon limits on the RF linearity are of a more fundamental nature and intrinsic to the technology. 
%intrinsic to the technology and are of a more fundamental nature. 
In the literature, RF linearity is often assumed to be closely linked to the device transconductance ($g_\mathrm{m}$), and the flatness of $g_\mathrm{m}$ with respect to drain current is frequently used as a proxy for RF linearity.~\cite{chenBellShapeGM2016, wuTransconductanceCollapse2005,russoSourceGateDistance2007,  joglekarVTLinearity2017, azadMultimetalGateGaN2023, tarakjiLargeSignalIII-N2003} Consequently, it has become common practice in the device community to seek transistor designs or biasing conditions where $g_\mathrm{m3}$, the second derivative of $g_\mathrm{m}$, is minimized in an attempt to achieve a linear GaN HEMT.~\cite{inoue2013linearity, tang2017simulation} However, $g_\mathrm{m}$ is fundamentally a small-signal metric, whereas RF linearity is more appropriately characterized using large-signal metrics such as output power, gain, power-added efficiency (PAE), output 1-dB compression point (OP$_\mathrm{1dB}$), and third-order output intercept power (OIP3).

This paper first examines the intrinsic limits imposed by LO phonons on the large-signal and RF linearity performance of a fabricated GaN HEMT, and subsequently investigates the correlation between $g_\mathrm{m}$ flatness and RF linearity. The analysis is carried out using a TCAD framework based on Fermi kinetics transport (FKT) that incorporates the full band structure of GaN and nonequilibrium phonon dynamics. This framework enables systematic analysis of the phonon bottleneck and shows that when LO phonon heating is considered, OP$_\mathrm{1dB}$ and OIP3 degrade by $\sim$3\,dB even for LO phonon lifetimes as short as 30\,fs in GaN HEMTs. Furthermore, the study demonstrates that optimizing $g_\mathrm{m3}$ does not necessarily eliminate large-signal nonlinear behavior.

\section{Device Setup and Simulation Details}
\subsection{Device Configuration}

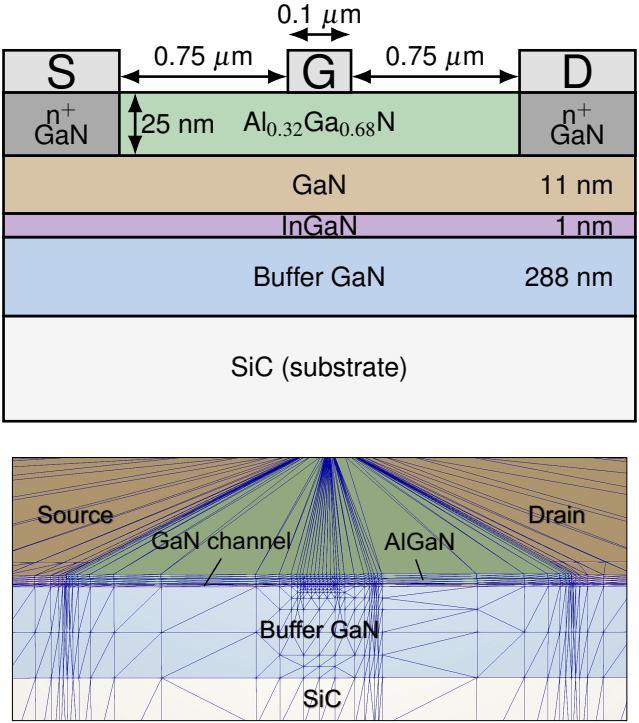
\begin{figure}[!h]
    \centering
    \begin{subfigure}{\linewidth}
    \centering
    \begin{tikzpicture}[
      font=\sffamily,
      >=Latex,
      dim/.style={Latex-Latex, line width=1pt},
      box/.style={draw=black, line width=1pt},
      scale=0.7, transform shape,
    ]
    
    % --------------------------
    % Geometry (arbitrary units, visually matched)
    % --------------------------
    \def\W{12.0}          % total width
    \def\xL{0}            % left edge
    \def\xR{\W}           % right edge
    
    % Layer thicknesses (visual units)
    \def\hSiC{2.0}
    \def\hBuf{1.5}
    \def\hIn{0.45}
    \def\hGaN{1.1}
    \def\hAl{1.2}
    \def\hTopMetal{0.8}
    \def\hNplus{1.4}      % n+ GaN thickness (for the 25 nm arrow region)
    
    % Y coordinates
    \def\y0{0}
    \pgfmathsetmacro{\ySiC}{\y0}
    \pgfmathsetmacro{\yBuf}{\ySiC+\hSiC}
    \pgfmathsetmacro{\yIn}{\yBuf+\hBuf}
    \pgfmathsetmacro{\yGaN}{\yIn+\hIn}
    \pgfmathsetmacro{\yAl}{\yGaN+\hGaN}
    \pgfmathsetmacro{\yTop}{\yAl+\hAl}
    
    % Contact / gate lateral sizes (visual units)
    \def\wContact{2.2}
    \def\wGate{1.2}
    
    \pgfmathsetmacro{\xS}{0.0}                       % source contact left
    \pgfmathsetmacro{\xSright}{\xS+\wContact}         % source contact right
    \pgfmathsetmacro{\xDright}{\xR}               % drain contact right
    \pgfmathsetmacro{\xD}{\xDright-\wContact}         % drain contact left
    
    % Place gate centered between S and D with spacing arrows like the figure
    \pgfmathsetmacro{\xG}{(\xSright+\xD)/2 - \wGate/2}
    \pgfmathsetmacro{\xGright}{\xG+\wGate}
    
    % --------------------------
    % Colors (approximate)
    % --------------------------
    \definecolor{cSiC}{RGB}{245,245,245}     % SiC – near white
    \definecolor{cBuf}{RGB}{190,210,235}     % Buffer GaN – soft blue
    \definecolor{cInGaN}{RGB}{200,175,215}   % InGaN – muted violet
    \definecolor{cGaN}{RGB}{215,195,165}     % GaN – warm beige
    \definecolor{cAlGaN}{RGB}{185,215,185}   % AlGaN – muted green
    \definecolor{cNplus}{RGB}{170,170,170}   % n+ GaN – neutral gray
    \definecolor{cMetal}{RGB}{225,225,225}   % Metal – light gray
    
    % --------------------------
    % Stack layers (full width)
    % --------------------------
    \draw[box,fill=cSiC] (\xL,\ySiC) rectangle (\xR,\yBuf);
    \node at ({(\xL+\xR)/2},{(\ySiC+\yBuf)/2}) {\Large SiC (substrate)};
    
    \draw[box,fill=cBuf] (\xL,\yBuf) rectangle (\xR,\yIn);
    \node at ({(\xL+\xR)/2},{(\yBuf+\yIn)/2}) {\Large Buffer GaN};
    \node[anchor=east] at (\xR-0.3,{(\yBuf+\yIn)/2}) {\Large 288 nm};
    
    \draw[box,fill=cInGaN] (\xL,\yIn) rectangle (\xR,\yGaN);
    \node at ({(\xL+\xR)/2},{(\yIn+\yGaN)/2}) {\Large InGaN};
    \node[anchor=east] at (\xR-0.3,{(\yIn+\yGaN)/2}) {\Large 1 nm};
    
    \draw[box,fill=cGaN] (\xL,\yGaN) rectangle (\xR,\yAl);
    \node at ({(\xL+\xR)/2},{(\yGaN+\yAl)/2}) {\Large GaN};
    \node[anchor=east] at (\xR-0.3,{(\yGaN+\yAl)/2}) {\Large 11 nm};
    
    % AlGaN barrier (full width top layer)
    \draw[box,fill=cAlGaN] (\xL,\yAl) rectangle (\xR,\yTop);
    \node at ({(\xL+\xR)/2},{(\yAl+\yTop)/2}) {\Large Al$_{0.32}$Ga$_{0.68}$N};
    \node[anchor=east] at (\xR-0.3,{(\yAl+\yTop)/2}) {\Large 13 nm};
    
    % --------------------------
    % n+ GaN regions under S/D (25 nm label arrow corresponds to this block height)
    % --------------------------
    % Make n+ GaN start exactly at the AlGaN bottom edge
    \pgfmathsetmacro{\yNbot}{\yAl}
    
    \draw[box,fill=cNplus] (\xS,\yNbot) rectangle (\xSright,\yTop);
    \node at ({(\xS+\xSright)/2},{(\yNbot+\yTop)/2}) {\shortstack{\Large n$^{+}$ \\[-1mm] \Large GaN}};
    
    \draw[box,fill=cNplus] (\xD,\yNbot) rectangle (\xDright,\yTop);
    \node at ({(\xD+\xDright)/2},{(\yNbot+\yTop)/2}) {\shortstack{\Large n$^{+}$\\[-1mm]\Large GaN}};

    % --------------------------
    % Metal blocks (S, G, D)
    % --------------------------
    \draw[box,fill=cMetal] (\xS,\yTop) rectangle (\xSright,\yTop+\hTopMetal);
    \node at ({(\xS+\xSright)/2},{\yTop+\hTopMetal/2}) {\Huge S};
    
    \draw[box,fill=cMetal] (\xG,\yTop) rectangle (\xGright,\yTop+\hTopMetal);
    \node at ({(\xG+\xGright)/2},{\yTop+\hTopMetal/2}) {\Huge G};
    
    \draw[box,fill=cMetal] (\xD,\yTop) rectangle (\xDright,\yTop+\hTopMetal);
    \node at ({(\xD+\xDright)/2},{\yTop+\hTopMetal/2}) {\Huge D};
    
    % --------------------------
    % Dimension arrows and labels
    % --------------------------
    
    % 0.1 um above gate
    \draw[dim] (\xG,\yTop+\hTopMetal+0.3) -- (\xGright,\yTop+\hTopMetal+0.3);
    \node[above] at ({(\xG+\xGright)/2},{\yTop+\hTopMetal+0.3}) {\Large 0.1 $\mu$m};
    
    % 0.75 um left spacing: from S right edge to G left edge
    \draw[dim] (\xSright,\yTop+0.3) -- (\xG,\yTop+0.3);
    \node[above] at ({(\xSright+\xG)/2},{\yTop+0.3}) {\Large 0.75 $\mu$m};
    
    % 0.75 um right spacing: from G right edge to D left edge
    \draw[dim] (\xGright,\yTop+0.3) -- (\xD,\yTop+0.3);
    \node[above] at ({(\xGright+\xD)/2},{\yTop+0.3}) {\Large 0.75 $\mu$m};
    
    % 25 nm vertical arrow at left (height of n+ GaN block)
    \draw[dim] ({\xSright+0.3},\yNbot) -- ({\xSright+0.3},\yTop);
    \node[right] at ({\xSright+0.3},{(\yNbot+\yTop)/2}) {\Large 25 nm};
    \end{tikzpicture}
    \end{subfigure}
    \begin{subfigure}{0.95\linewidth}
    \centering
    \includegraphics[width=\linewidth]{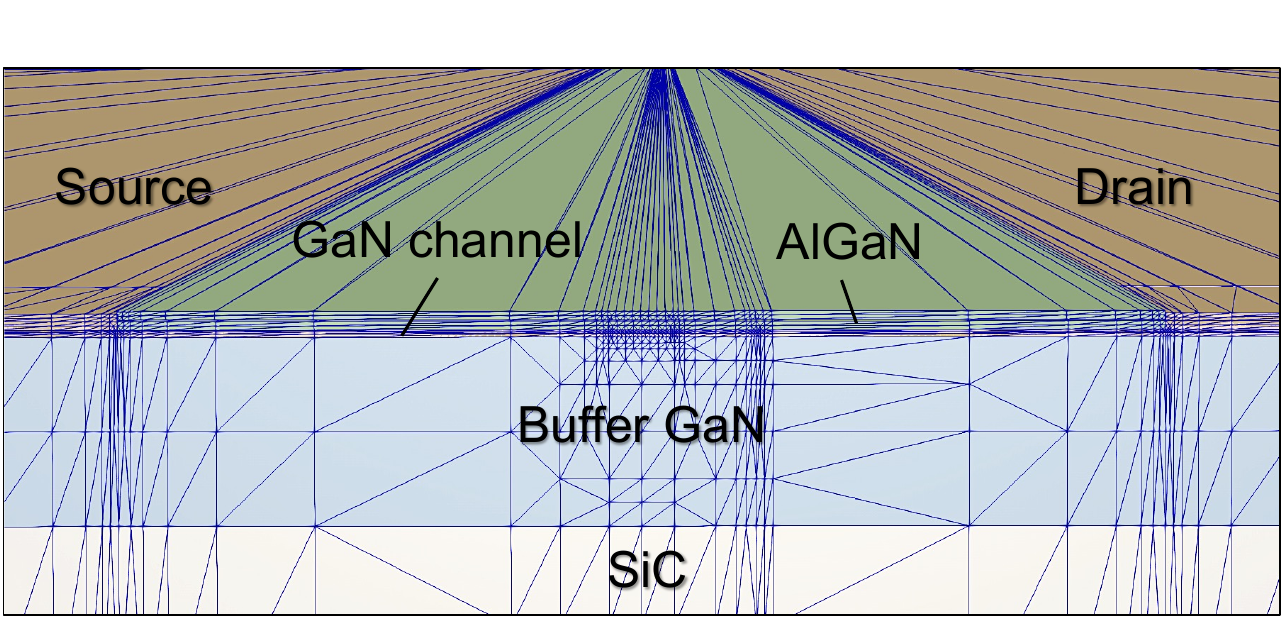}
    \end{subfigure}
    \caption{Schematic cross section of the AlGaN/GaN HEMT (top) and the meshed structure (bottom). Gate Schottky barrier height: 1.2\,eV; polarization sheet charge: $1.115\times 10^{13}$\,cm$^{-2}$; surface  trap density in the source (drain) access region: $1.8\times 10^{12}$\,cm$^{-2}$ ($5.2\times 10^{12}$\,cm$^{-2}$); acoustic phonon deformation potential: 8.3\,eV; LO phonon energy: 92\,meV; LO phonon deformation potential: $10^9$\,eV/cm.}
    \label{fig:device_sch}
\end{figure}

The device schematic is presented in Fig.~\ref{fig:device_sch}. The device dimensions are based on the GaN HEMT device grown using metal-organic chemical vapor deposition (MOCVD) method,~\cite{palaciosEmodeGaN2006} consisting of a GaN buffer grown on a SiC substrate for improved heat removal, followed by a GaN channel and an $\mathrm{Al}_{0.32}\mathrm{Ga}_{0.64}\mathrm{N}$ barrier layer that induces a high-density two-dimensional electron gas (2DEG) at the interface through spontaneous and piezoelectric polarization. A thin layer of $\mathrm{{In}_{0.1}}\mathrm{{Ga}_{0.9}}\mathrm{N}$ was grown on the GaN buffer for better electron confinement. Source and drain regions are heavily doped to form low-resistance ohmic contacts, while the gate contact forms a Schottky barrier to the recessed AlGaN barrier, providing electrostatic control of the channel. For the thermal boundary conditions, the bottom surface of the substrate is held at a fixed temperature of 300\,K using the Dirichlet boundary condition, while all remaining exposed surfaces are assigned Neumann boundary conditions. The perfect electric conductor (PEC) regions at the top of the device are assumed to be isothermal.

\vspace{-10pt}
\subsection{Fermi Kinetics Transport}
Fermi kinetics transport is a deterministic Boltzmann transport solver, in which mobile carriers are treated as interacting ideal Fermi gases.~\cite{grupenHeatFlowTransport2009, grupenEnergyTransportGaAs2011, grupenFullWaveGaN2016}
%{\color{black} Unlike drift--diffusion and conventional hydrodynamic models, FKT does not rely on empirical field-dependent mobilities, electron thermal conductivities, or adjustable energy relaxation parameters. Rather, the}
The defining feature of the model is its treatment of electronic heat flow: instead of invoking Fourier’s law, heat transfer is computed directly from a thermodynamic identity for ideal Fermi gases, ensuring that particle and energy exchange between local carrier populations obeys the second law of thermodynamics. It also gives FKT device simulations improved numerical stability and greater flexibility.~\cite{miller2018computational, tungaTCADComparison2022, miller2023recent, white2023large} For example, instead of assuming parabolic bands and empirical field-dependent mobilities, FKT can incorporate realistic electronic band structure and energy-dependent mobilities derived directly from quantum mechanical scattering rates.~\cite{grupenFullWaveGaN2016}

For the simulations considered here, FKT carrier dynamics are coupled to quasi-static fields by self-consistently solving Poisson’s equation together with mobile electron particle and energy continuity, as well as lattice energy conservation equations. Nonequilibrium LO-phonon dynamics are treated explicitly through the following rate equation:~\cite{dastiderFullBandGaN2026}
\begin{equation}
\frac{1}{8\pi^3} \int d{\bf{q}} \frac{dn_\mathrm{q}}{dt}  = U_\mathrm{LO} -\frac{1}{8\pi^3} \int d {\bf{q}}\frac{n_\mathrm{q} - n_\mathrm{q}^0}{\tau_{\mathrm{LO}}},
\label{eqn:LO_decay}
\end{equation}

\noindent where $n_\mathrm{q}$ is the phonon occupation number, and $n_\mathrm{q}^0$ is its value when LO phonon temperature equals the acoustic phonon temperature $T_{\rm LO}=T_{\rm A}$. \textcolor{black}{Acoustic phonon branches are represented through a single effective acoustic phonon temperature, $T_\mathrm{A}$. $U_\mathrm{LO}$, the LO phonon collision operator, is the sum of deformation-potential and polar optical phonon contributions,}

\vspace{-15pt}

\begin{equation}
\color{black}
\begin{aligned}
U_{\mathrm{LO}} &= U_{\mathrm{odp}} + U_{\mathrm{pop}} \\
U_{\mathrm{odp(pop)}} &=
\int_{E_i}
\Biggl[
\int_{\bm{k_i}}
\rho_{\bm{k}}(E)
\Biggl(
\int_{k_f}
W_{\bm{k,k'}}^{\mathrm{odp(pop)}} \,\rho_{\bm{k'}}(E-\hbar\omega)\, d\bm{k'} \, d\bm{k}
\Biggr)
\Biggr] \\
&\quad \times
\Bigl[
(n_q+1)f_i(1-f_f)-n_q f_f(1-f_i)
\Bigr]\, dE,
\end{aligned}
\end{equation}

\noindent \textcolor{black}{where $W_{\textit{\textbf{k,k'}}}$ is the scattering probability between the initial and final $\textit{\textbf{k}}$ states, as obtained from Fermi's golden rule, $\hbar\omega$ is phonon energy, $\rho_{\textbf{k}}$ is the local electron density of states, $f_i$ is the Fermi--Dirac distribution function. In this work, ``hot LO phonons'' refers to an nonequilibrium excess LO-phonon population represented by an equivalent $T_{\textrm{LO}}$ that is elevated above $T_{\mathrm{A}}$. The LO-phonon occupation $n_\mathrm{q}$ used in the electron--LO-phonon scattering rates is evaluated using the Bose--Einstein distribution at $T_{\mathrm{LO}}$. Instead of solving a fully mode-resolved distribution $n(q)$ for each LO phonon wave vector, the LO-phonon population is evolved using the rate equation, Eq.~(\ref{eqn:LO_decay}).~\cite{dastiderFullBandGaN2026}} The rate equation balances electron-induced phonon generation against finite-lifetime decay into acoustic modes, with the resulting hot-phonon population feeding back into the electron scattering rates. \textcolor{black}{The LO phonons are approximated as stationary because their small group velocity, together with the ultrashort lifetimes considered here, leads to a negligible propagation length relative to the characteristic device dimensions.} The inclusion of full band structure, physically derived scattering mechanisms from Fermi's golden rule, and the inclusion of LO phonon dynamics~\cite{dastiderFullBandGaN2026} allow FKT to capture hot-electron and hot-phonon effects with high accuracy and computational efficiency.~\cite{tungaTCADComparison2022}

\vspace{-10pt}
\subsection{Large-Signal Setup}
For the large-signal analysis, a transient physics-based simulation is performed, followed by frequency-domain post-processing. In this approach, the nonlinear semiconductor transport equations are solved directly in the time domain under periodic excitation, and the resulting terminal waveforms are subsequently decomposed into harmonic components to extract RF performance metrics.

For each bias condition, the simulator generates time-dependent terminal voltages and currents at the input and output ports, denoted as $v_\mathrm{in}(t)$, $i_\mathrm{in}(t)$, $v_\mathrm{out}(t)$, and $i_\mathrm{out}(t)$. These waveforms capture nonlinear effects, including waveform distortion and harmonic generation. To ensure the device has reached a stable state, the analysis identifies a Continuous Wave (CW) window at the end of the transient simulation. Therefore, 
% \sout{To ensure periodic steady-state {\color{black} do you mean continuous wave (CW)?} operation,}
only the final excitation cycle, \emph{i.e.,} the time window $t \in \left[T_{\mathrm{end}}-\frac{1}{f_0},\,T_{\mathrm{end}}\right] $
is selected, where $f_\mathrm{0}=10$\,GHz is the fundamental driving frequency. Harmonic decomposition is performed using a discrete Fourier transform of the steady-state waveforms. The complex voltage and current harmonics are obtained from fast Fourier transforms (FFTs):
\begin{equation}
V^{(k)} = \frac{2}{N}\,\mathrm{FFT}\{v(t)\}, \qquad
I^{(k)} = \frac{2}{N}\,\mathrm{FFT}\{i(t)\},
\end{equation}
where $k$ denotes harmonic order and $N$ is the number of time samples. Retaining the first several harmonics provides a compact frequency-domain representation of the nonlinear terminal behavior.

Large-signal power quantities are computed using the power-wave formalism. For a reference impedance $Z_0$, the incident and reflected waves at each port are defined as
\begin{equation}
a = \frac{v + Z_0 i}{2\sqrt{Z_0}}, \qquad
b = \frac{v - Z_0 i}{2\sqrt{Z_0}} ,
\end{equation}
where $Z_\mathrm{0} = 50~\Omega$. From these quantities, nonlinear reflection coefficients and effective impedances are evaluated, enabling direct comparison with measurement-based RF metrics.

Delivered input and output powers are obtained from the power-wave differences,
\begin{equation}
P_{\mathrm{in,del}}=\frac{|a_1|^2-|b_1|^2}{2}, \qquad
P_{\mathrm{out}}=\frac{|b_2|^2-|a_2|^2}{2},
\end{equation}
The available input power, \(P_{\mathrm{in,avail}}\) (denoted hereafter as \(P_{\mathrm{in}}\)), is calculated from the $P_\mathrm{in,del}$ after accounting for source--input mismatch according to
\begin{equation}
P_{\mathrm{in,avail}} =
\frac{P_{\mathrm{in,del}}}
{1-\left|
\frac{Z_{\mathrm{in}}-Z_\mathrm{S}^{*}}
{Z_{\mathrm{in}}+Z_\mathrm{S}^{*}}
\right|^{2}},
\end{equation}
where \(Z_{\mathrm{in}}\) is the input impedance obtained from the simulated terminal waves and \(Z_\mathrm{S}\) is the source impedance. DC power consumption is extracted from the zero-frequency components of the terminal quantities. These values are used to compute standard large-signal RF metrics, including gain, PAE, OP$_\mathrm{1dB}$, OIP3, \emph{etc.}

\vspace{-15pt}
\section{Results and Discussions}
\vspace{-5pt}
\subsection{Device Simulation} \label{sec:dev_sim}

\vspace{-10pt}

\begin{figure}[h!]
    \centering
    \includegraphics[width=0.8\linewidth]{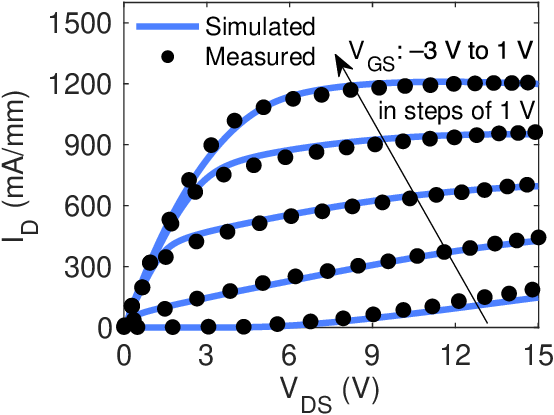}
    %\vspace{-15pt}
    \caption{Drain current $I_{\rm D}$ versus drain voltage $V_{\rm DS}$ with different gate biases $V_{\rm GS}$ for the GaN HEMT structure in Fig.~\ref{fig:device_sch}. Measurements reported by Marino {\em et al.}~\cite{marinoDislocationsGaNHEMT2010}}
    \label{fig:I-V}
\end{figure}

\begin{figure}[h!]
\vspace{-15pt}
    \centering
    \includegraphics[width=0.75\linewidth]{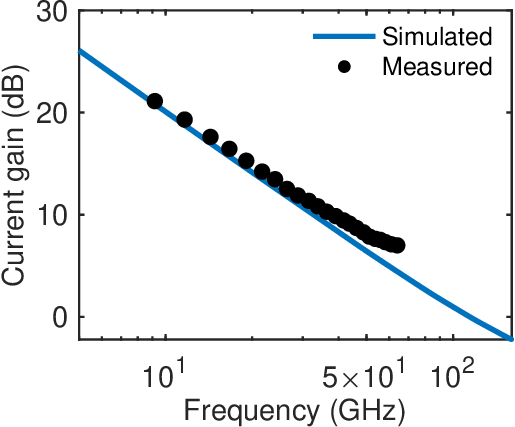}
    %\vspace{-15pt}
    \caption{Small signal current gain versus frequency for the GaN HEMT in Fig.~\ref{fig:device_sch}. Measured data reported by Palacios {\em et al.}~\cite{palaciosEmodeGaN2006}}
    \label{fig:h21}
\end{figure}

The simulated output characteristics and the small-signal current gain of the GaN HEMT are compared with the experimental measurements in Figs.~\ref{fig:I-V} and \ref{fig:h21}, respectively. Device parameters are noted in the caption of Fig.~\ref{fig:device_sch}. The results demonstrate excellent agreement across the examined bias range with both acoustic and LO phonon heating included in the model.

\vspace{-5pt}

\begin{figure*}
    \centering
    \includegraphics[width=0.9\linewidth]{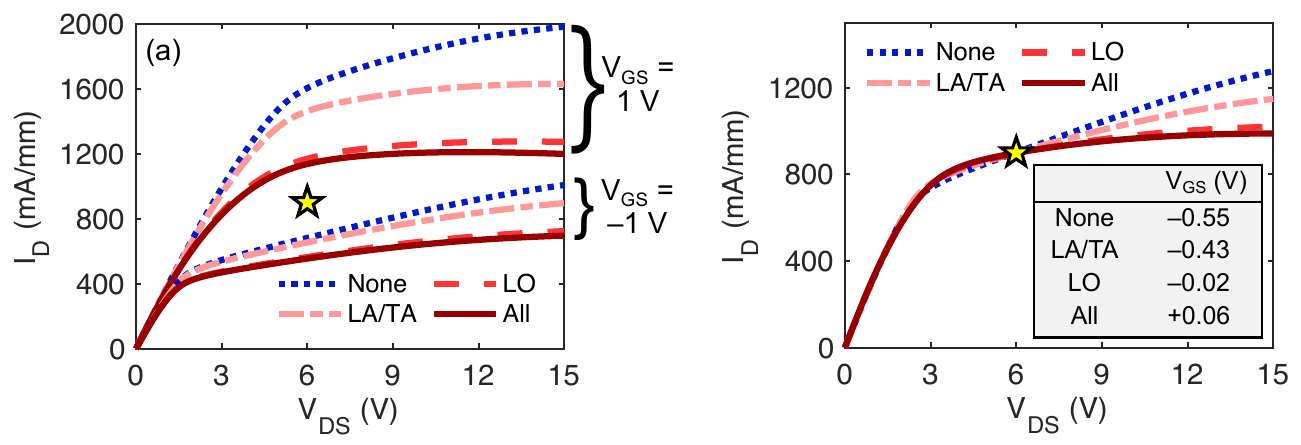}
    %\vspace{-15pt}
    \caption{(a) The simulated steady-state current at $V_\mathrm{GS} = 1$\,V and $V_\mathrm{GS} = -1$\,V highlights the impact of LO phonon heating on DC response under high bias conditions. %The steady-state current under four heating conditions reaching the targeted Q-point.
    The symbol marks the quiescent (Q)-point: $I_\mathrm{D} = 900$\,mA/mm, $V_\mathrm{DS} = 6$\,V. (b) Output curves at the Q-point for various cases of phonon heating. For each case, the gate bias is adjusted to arrive at the Q point as indicated in the inset table.  
    %Incorporating LO phonon heating leads to a substantial reduction in current compared to models that include only acoustic phonon heating (LA/TA) or neglect phonon heating entirely. Notably, even without accounting for acoustic phonon heating, the LO phonon contribution alone is sufficient to significantly degrade the output current.
    }
    \label{fig:idvd_Q}
\end{figure*}

\vspace{5pt}

The LO phonon lifetime is chosen to be 30\,fs, which is consistent with experimental studies on AlGaN/GaN heterostructures~\cite{matulionisPlasmonHeat2009, liberisHotPhononLifetime2014} and aligns with our prior work, which established an upper bound of $\tau_\mathrm{LO} \leq 40$\,fs for GaN HEMTs.~\cite{dastiderFullBandGaN2026}

% It is to be noted that LO phonon lifetime ($\tau_{\mathrm{LO}}$) plays a significant role in the DC characteristics~\cite{b23}. In the literature, $\tau_{\mathrm{LO}}$ varies widely depending on different measurements \cite{b20,b21,b22}. Based on our earlier work~\cite{b23}, we are using a value of 30\,fs in this study, which is supported by previous experimental studies on AlGaN/GaN heterostructures~\cite{b20,b21}.

% In the context of nonlinearity, RF performance metrics serve as key indicators, providing practical insight into the impact of various heating mechanisms. Before analyzing the large-signal RF behavior, 
%\vspace{-15pt}

\vspace{-15pt}
\subsection{Large-Signal RF Metrics}
\vspace{-10pt}

We examine the steady-state output characteristics under four different phonon heating conditions in Fig.~\ref{fig:idvd_Q}(a). The case labeled as ``All'' includes both acoustic and LO phonon heating, ``LO'' includes only LO phonon heating while acoustic phonon temperature is set equal to the ambient temperature, ``LA/TA'' considers longitudinal+transverse acoustic phonon heating but ignores LO phonon heating (\emph{i.e.} LO phonon temperature is the same as acoustic phonon temperature), whereas ``None'' ignores all types of phonon heating. 
The inclusion of LO phonon heating impacts the current significantly at higher gate and drain biases, as shown in Fig.~\ref{fig:idvd_Q}(a). The quiescent (Q) point for each large-signal simulation is marked by the star symbol in Fig.~\ref{fig:idvd_Q}(a) and corresponds to a different gate voltage for each of the four heating conditions, as shown in Fig.~\ref{fig:idvd_Q}(b).   
For additional commentary on the choice of the quiescent bias point, please consult the Appendix.

\begin{figure}[h!]
\vspace{-5pt}
    \centering
    \includegraphics[width=0.79\linewidth]{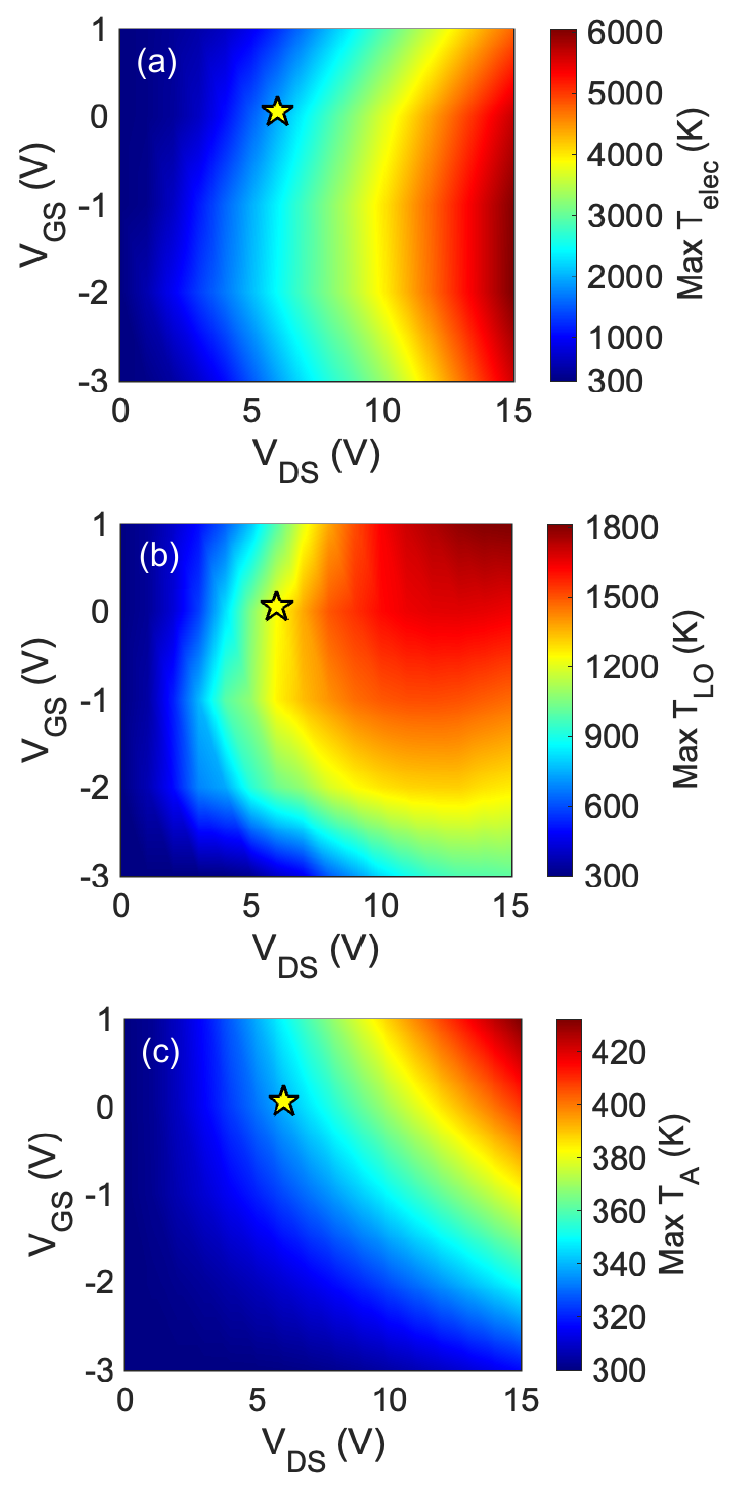}
    \vspace{-5pt}
    \caption{Two-dimensional heat maps of the peak (a) electron, (b) LO phonon, and (c) acoustic phonon temperatures in the GaN channel. The symbol indicates the Q-point.}
    \label{fig:2Dtemp}
\end{figure}

The large-signal response of the HEMT is simulated at the Q-point for each of the four phonon heating cases. The electron and phonon temperatures at the Q-point are examined in Fig.~\ref{fig:2Dtemp} and it is found that for $\tau_\mathrm{LO}=30$\,fs in the ``All'' case, the peak electron temperature reaches 1943\,K and tracks the LO phonon temperature which reaches up to 1250\,K, while acoustic phonon temperature rises to 340\,K.
Further increasing the drain bias at a given gate bias consistently increases the temperature of electrons and phonons.

\begin{figure*}
    \centering
    \includegraphics[width=0.65\linewidth]{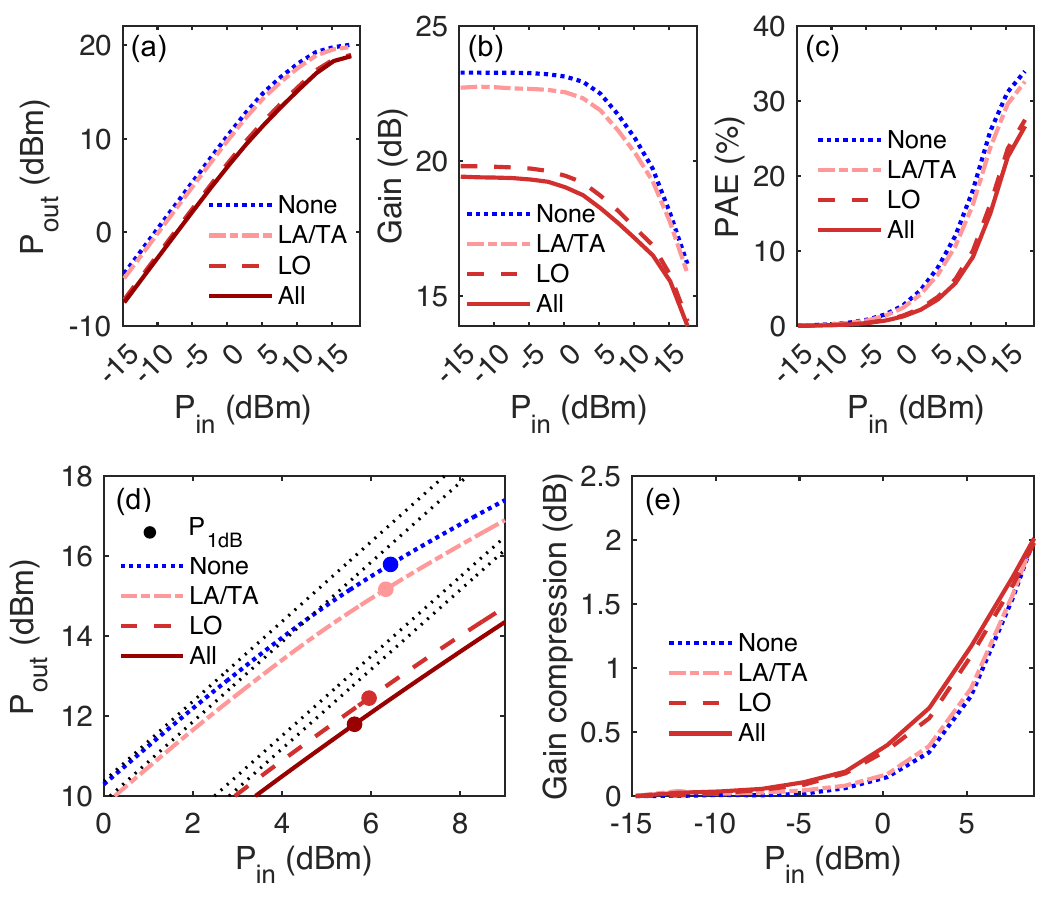}
    %\vspace{-15pt}
    \caption{Large-signal analysis of the GaN HEMT showing the impact of LO phonon heating. (a) Output power, $P_\mathrm{out}$ versus available input power, $P_\mathrm{in}$, (b) gain versus $P_\mathrm{in}$, (c) power-added efficiency (PAE) versus $P_\mathrm{in}$, (d) $P_\mathrm{out}$-$P_\mathrm{in}$ curve showing the 1-dB compression point, and (e) gain compression versus $P_\mathrm{in}$. Gain is defined as $P_{\mathrm{out}}/P_{\mathrm{in,del}}$, and PAE is defined as $(P_{\mathrm{out}} - P_{\mathrm{in,del}})/P_{\mathrm{DC}}$. The Q-point and other parameters: $I_\mathrm{D} = 900$\,mA/mm, $V_\mathrm{DS} = 6$\,V, $f_\mathrm{0} = 10$\,GHz, $Z_\mathrm{L} = 70 + j6.0~\Omega$.}
    \label{fig:PoutPin_Gain_PAE_OP1dB}
\end{figure*}

As shown in Fig.~\ref{fig:PoutPin_Gain_PAE_OP1dB}(a), at a given $P_\mathrm{in}$, there is a consistent reduction in $P_\mathrm{out}$ as phonon heating is turned on, with significant degradation occurring due to LO phonon heating despite their ultrafast decay into acoustic modes. Likewise, the gain and PAE of the HEMT, plotted in Figs.~\ref{fig:PoutPin_Gain_PAE_OP1dB}(b) \& (c), respectively, show a systematic degradation with phonon heating where only LA/TA phonon heating is less severe whereas LO phonon heating causes a much more dramatic degradation in the large-signal metrics of the HEMT. 
The time-domain waveforms together with the corresponding dynamic load lines, which provide additional insight into $P_\mathrm{out}-P_\mathrm{in}$ mapping and the nonlinear output behavior, are presented in the Appendix.

Figure~\ref{fig:PoutPin_Gain_PAE_OP1dB}(d) shows the 1-dB compression points for all four heating scenarios. The OP$_\mathrm{1dB}$ decreases slightly from 15.79\,dBm to 15.17\,dBm when acoustic phonon heating is enabled, while LO phonon heating remains disabled. In contrast, enabling LO phonon heating while suppressing acoustic heating leads to a much more pronounced degradation, reducing OP$_\mathrm{1dB}$ to 12.45\,dBm. When both LO and acoustic phonon heating are included, OP$_\mathrm{1dB}$ further decreases to 11.79\,dBm. Figure~\ref{fig:PoutPin_Gain_PAE_OP1dB}(e) provides a clearer view of the gain compression behavior, showing that inclusion of LO phonon heating leads to an earlier onset and a more gradual increase in compression. This indicates softer gain compression in the cases with LO phonon heating compared to the sharper transition observed in the cases without LO phonon heating. These results highlight that LO phonons significantly impact GaN HEMT performance despite their ultrashort lifetimes of only a few tens of femtoseconds. 

\begin{figure}[h!]
    \centering
    \includegraphics[width=1.0\linewidth]{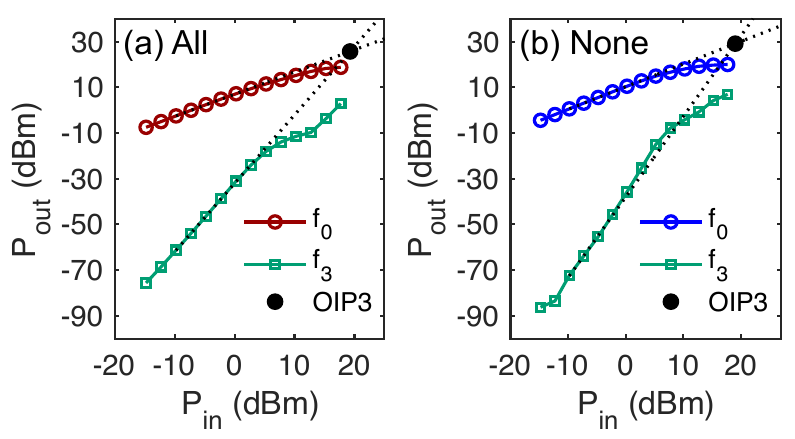}
    %\vspace{-15pt}
    \caption{OIP3 at a fixed Q-point ($V_\mathrm{DS}$ = 6\,V, $I_\mathrm{D}$ = 900\,mA/mm) for (a) ``All'' and (b) ``None'' heating conditions.}
    \label{fig:OIP3_none+all}
\end{figure}

The output power at third order intercept (OIP3) is an extrapolated metric that defines the output power of the amplifier where the power contained in the fundamental signal and the third harmonic at the output become equal.~\cite{alim2021experimental} A device with higher nonlinearity exhibits a lower value of OIP3.~\cite{shrestha2020high} From the large-signal time-domain simulations, we extract OIP3 for all cases of phonon heating (results for ``All'' and ``None'' are shown in Fig.~\ref{fig:OIP3_none+all}) and note a consistent trend: In the absence of all phonon heating the device exhibits an OIP3 of 29.15\,dBm, which degrades by $\sim 3$\,dB when LO phonon heating is turned on and by another 0.6\,dB when both LO and acoustic phonon heating are on.  

\begin{figure}[h!]
    \centering
    \includegraphics[width=0.75\linewidth]{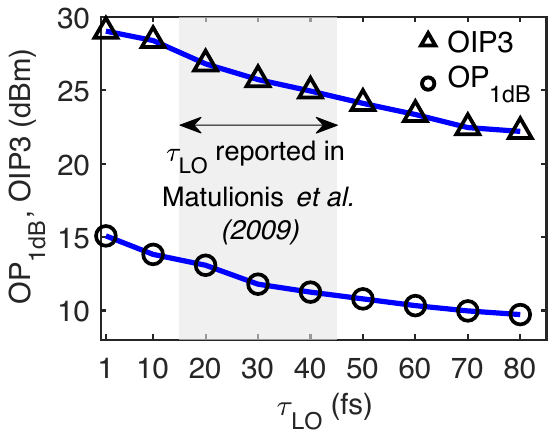}
    %\vspace{-15pt}
    \caption{Degradation of OP$_\mathrm{1dB}$ and OIP3 at a fixed Q-point ($V_\mathrm{DS}$ = 6\,V, $I_\mathrm{D}$ = 900\,mA/mm) with increasing LO phonon lifetime. The shaded region corresponds to experimentally determined $\tau_\mathrm{LO}$ values for an AlGaN/GaN heterostructure.~\cite{matulionisPlasmonHeat2009}}
    \label{fig:OIP3_OP1dB_tauLO}
    \vspace{-10pt}
\end{figure}

The variation of OP$_\mathrm{1dB}$ and OIP3 with $\tau_\mathrm{LO}$, shown in Fig.~\ref{fig:OIP3_OP1dB_tauLO}, further highlights the negative impact of LO phonon heating on the RF metrics. As $\tau_\mathrm{LO}$ increases from 1\,fs to 80\,fs, both OP$_\mathrm{1dB}$ and OIP3 degrade by $\sim (5-6)$\,dB.   
The shaded region in the figure highlights the range of $\tau_\mathrm{LO}$ reported by Matulionis \textit{et al.} in an experimental AlGaN/GaN heterostructure,~\cite{matulionisPlasmonHeat2009} which aligns with our group's earlier findings.~\cite{dastiderFullBandGaN2026} A summary of the impact of phonon heating on various large-signal RF metrics of the device at $\tau_\mathrm{LO}=30\,$fs is presented in Table~\ref{tab:performance_op1db}.

\begin{table}[h!]
\vspace{-5pt}
\caption{RF performance metrics of the GaN HEMT obtained for various phonon heating conditions. $\tau_\mathrm{LO}=30$\,fs for ``All'' and ``LO'' cases.}
\label{tab:performance_op1db}
\centering
\renewcommand{\arraystretch}{1.2}
\begin{tabular}{lcccc}
\hline
Configuration 
& OP$_{\textrm{1dB}}$ 
& OIP3 
& Gain @OP$_{\textrm{1dB}}$ 
& PAE @OP$_{\textrm{1dB}}$ \\
& (dBm) & (dBm) & (dB) & (\%) \\
\hline
None  & 15.79 & 29.15 & 22.16 & 9.68 \\
LA/TA & 15.17 & 28.20 & 21.59 & 8.37 \\
LO    & 12.45 & 26.29 & 18.56 & 4.47 \\
All   & 11.79 & 25.73 & 18.17 & 3.84 \\
\hline
\end{tabular}
\end{table}

\vspace{-10pt}
\subsection{RF Nonlinearity and Transconductance Flatness}
\vspace{-15pt}

\begin{figure}
    \centering
    \includegraphics[width=0.8\linewidth]{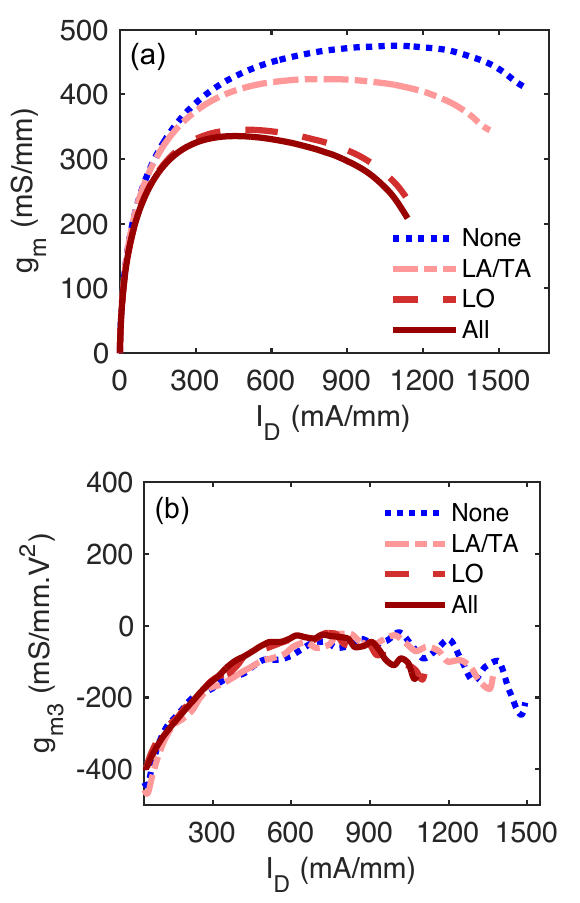}
    %\vspace{-10pt}
    \caption{Impact of phonon heating on (a) transconductance ($g_\mathrm{m}$) and its higher-order derivative, (b) $g_\mathrm{m3}$ at $V_\mathrm{DS}$ = 6\,V, omitting the subthreshold regime.}
    %{\color{black} could you please put horizontal dotted line for $g_{m2}=0$? the way it's plotted now, it's a little difficult to compare $g_{m2}$ magnitudes at $I_D=900$.}}
    \label{fig:gm_gm3}
    \vspace{-5pt}
\end{figure}

To investigate the relationship between $g_\mathrm{m}$ flatness and large-signal RF nonlinearity, we first plot $g_\mathrm{m}$ and its \textcolor{black}{second}-order derivative $g_\mathrm{m3}$ as a function of drain current at $V_\mathrm{DS}=6$\,V for different phonon-heating conditions. As expected, Fig.~\ref{fig:gm_gm3} indicates $g_\mathrm{m}$ improves by 41\% as phonon heating effects are progressively removed.

In contrast, at the chosen Q-point of large-signal analysis, $g_\mathrm{m3}$, which characterizes the flatness of the $g_\mathrm{m}$ curve, is only weakly sensitive to phonon heating, remaining within the range of 46--54\,mS/mm$\cdot$V$^{2}$ across the different cases. If $g_\mathrm{m}$ flatness were a reliable proxy for RF linearity, these results would suggest similar RF performance under all phonon-heating conditions. However, the large-signal simulations, summarized in Table~\ref
{tab:performance_op1db}, indicate otherwise.

\begin{figure}[h!]
%\vspace{-5pt}
    \centering
    \includegraphics[width=0.8\linewidth]{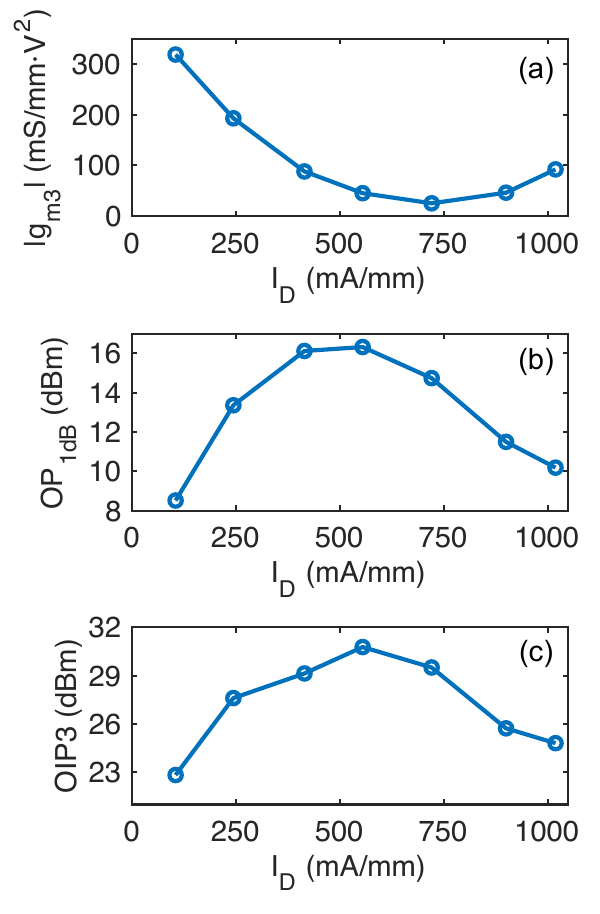}
    %\vspace{-15pt}
   % \vspace{-10pt}
    \caption{(a) $g_\mathrm{m3}$, (b) OP$_\textrm{1dB}$, and (c) OIP3 vs. drain current, $I_\textrm{D}$, at $V_\mathrm{DS}=6$\,V for ``All'' phonon heating condition.}
    \label{fig:gm3_RF}
\end{figure}

A complementary perspective on the relationship between $g_\mathrm{m}$ flatness and RF linearity is obtained by plotting $|g_\mathrm{m3}|$, OP$_\mathrm{1dB}$, and OIP3 as a function of drain current for the ``All'' case in Fig.~\ref{fig:gm3_RF}. Although $|g_\mathrm{m3}|$ gradually decreases and reaches its minimum magnitude (\emph{i.e.}, closest to zero) near $I_\mathrm{D}=723$\,mA/mm ($V_\mathrm{GS}=-0.5$\,V), the best RF performance is instead obtained at approximately $I_\mathrm{D}=555$\,mA/mm ($V_\mathrm{GS}=-1$\,V). This discrepancy further underscores the limited predictive capability of $g_\mathrm{m3}$ for large-signal RF linearity. The quantitative values corresponding to Fig.~\ref{fig:gm3_RF} are summarized in Table~\ref{tab:gm3_RF_all}. Notably, at $V_\mathrm{GS}=-1$\,V and $0$\,V, the extracted $g_\mathrm{m3}$ values are nearly identical, yet the resulting OP$_\mathrm{1dB}$ and OIP3 differ by 4.82\,dB and 5\,dB, respectively. These results clearly demonstrate that a small $|g_\mathrm{m3}|$ does not necessarily translate into improved RF linearity.

\begin{table}[h!]
\caption{Comparison of $g_\mathrm{m3}$ and large-signal RF metrics at $V_\mathrm{DS}=6$\,V for ``All'' phonon heating condition.}
\label{tab:gm3_RF_all}
\centering
\renewcommand{\arraystretch}{1.25}
\begin{tabular}{lcccc}
\hline
$V_\mathrm{GS}$ (V)
& $I_\mathrm{D}$ 
& $|g_\mathrm{m3}|$ 
& OP$_{\textrm{1dB}}$ 
& OIP3 \\
& (mA/mm) & (mS/mV$^2$) & (dBm) & (dBm) \\
\hline
$-$2.45 & 105  & 319 & 8.52  & 22.83 \\
$-$1.96 & 244  & 193 & 13.36 & 27.61 \\
$-$1.45 & 415  & 88  & 16.12 & 29.14 \\
\textbf{$-$1.02} & \textbf{555}  & \textbf{45}  & \textbf{16.32} & \textbf{30.78} \\
\textbf{$-$0.52} & \textbf{721}  & \textbf{25}  & \textbf{14.74} & \textbf{29.51} \\
\textbf{$+$0.06} & \textbf{900}  & \textbf{46}  & \textbf{11.79} & \textbf{25.73} \\
$+$0.50 & 1019 & 92  & 10.19 & 24.81 \\
\hline
\end{tabular}
\end{table}

\vspace{-15pt}
\section{Discussion}
\vspace{-10pt}

\begin{figure*}
    \centering
    \includegraphics[width=0.65\linewidth]{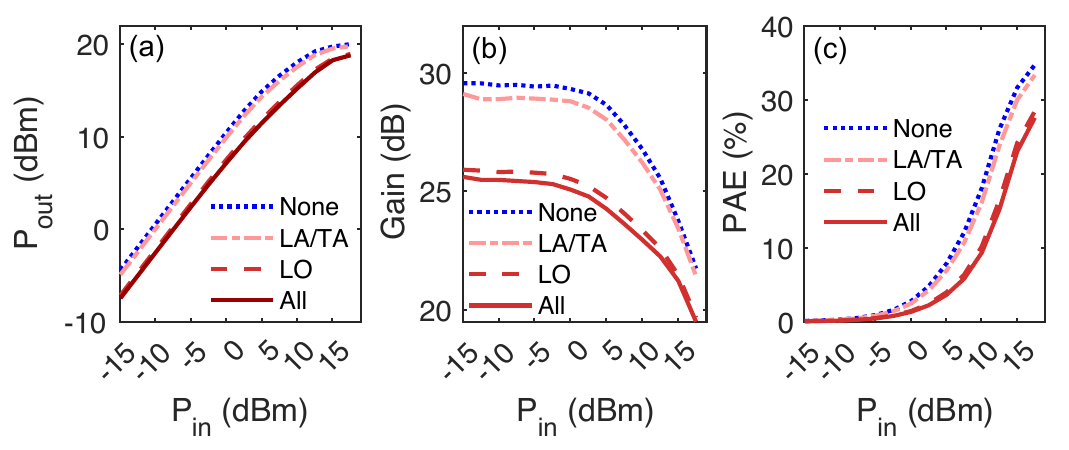}
    %\vspace{-15pt}
    \caption{Large-signal analysis of the GaN HEMT showing the impact of LO phonon heating at $f_\mathrm{0} = 5$\,GHz. (a) $P_\mathrm{out}$ versus $P_\mathrm{in}$, (b) gain versus $P_\mathrm{in}$, (c) PAE versus $P_\mathrm{in}$. The Q-point and other parameters are the same as in Fig.~\ref{fig:PoutPin_Gain_PAE_OP1dB}.}
    \label{fig:PoutPin_Gain_PAE_OP1dB_5GHz}
\end{figure*}

\textcolor{black}{The central objective of this work is to examine the intrinsic impact of LO-phonon heating on the large-signal RF performance of GaN HEMTs and to clarify the relationship between $g_\mathrm{m}$-based metrics and RF linearity. The quantitative impact of LO-phonon heating may depend on device geometry, including gate length, access-region dimensions, and field-plate design, through changes in the electric-field profile and heat generation. However, such changes are expected to affect the numerical values of current, temperature, OP$_\mathrm{1dB}$, OIP3, and PAE rather than the central conclusions: ultrafast LO phonons can produce significant out of equilibrium LO phonons which affect the device RF performance, including its RF linearity, while $g_\mathrm{m3}$ alone is not a reliable proxy for RF linearity. A systematic device-design study will be conducted in future work.}

\textcolor{black}{Similarly, the thermal boundary conditions may influence the quantitative acoustic-phonon temperature rise and the spatial thermal gradients in the device. However, these effects are not expected to alter the central conclusions, since the RF degradation reported here is dominated by LO-phonon heating rather than acoustic phonon heating.}

\textcolor{black}{The frequency dependence of LO-phonon-induced degradation is also relevant to the interpretation of the results. To examine the lower-frequency response, the large-signal analysis is repeated at 5~GHz, as shown in Fig.~\ref{fig:PoutPin_Gain_PAE_OP1dB_5GHz}. The $P_\mathrm{out}$--$P_\mathrm{in}$ and PAE characteristics remain unchanged from the 10\,GHz case, and therefore the extracted OP$_\mathrm{1dB}$, OIP3, and PAE at the 1-dB compression point are also the same. The gain increases by approximately 6\,dB at 5\,GHz, consistent with the frequency dependence of the transistor response. These results indicate that the relative degradation due to LO-phonon heating remains evident in the lower-frequency regime.}

\textcolor{black}{The behavior at higher frequencies is expected to be more complex. LO-phonon heating is expected to continue degrading RF performance, but this degradation will compete with other deleterious high-frequency effects. It is noted from Fig.~\ref{fig:h21} that at high frequencies ($\geq$38~GHz), the simulated results obtained under the quasi-static approximation deviate from the measured data. In this regime, inductive coupling between the metal contacts becomes pronounced.~\cite{grupenFullWaveGaN2016} This coupling is a propagating-wave effect governed by the rotational electric and magnetic fields in Ampere's and Faraday's laws. Finite electron momentum relaxation rates also become relevant to the high-frequency response. A full-wave electromagnetic treatment would therefore be required for more accurate modeling in this frequency range, which will be explored in future work.}

\begin{figure*}[t!]
    \centering
    \includegraphics[width=1\linewidth]{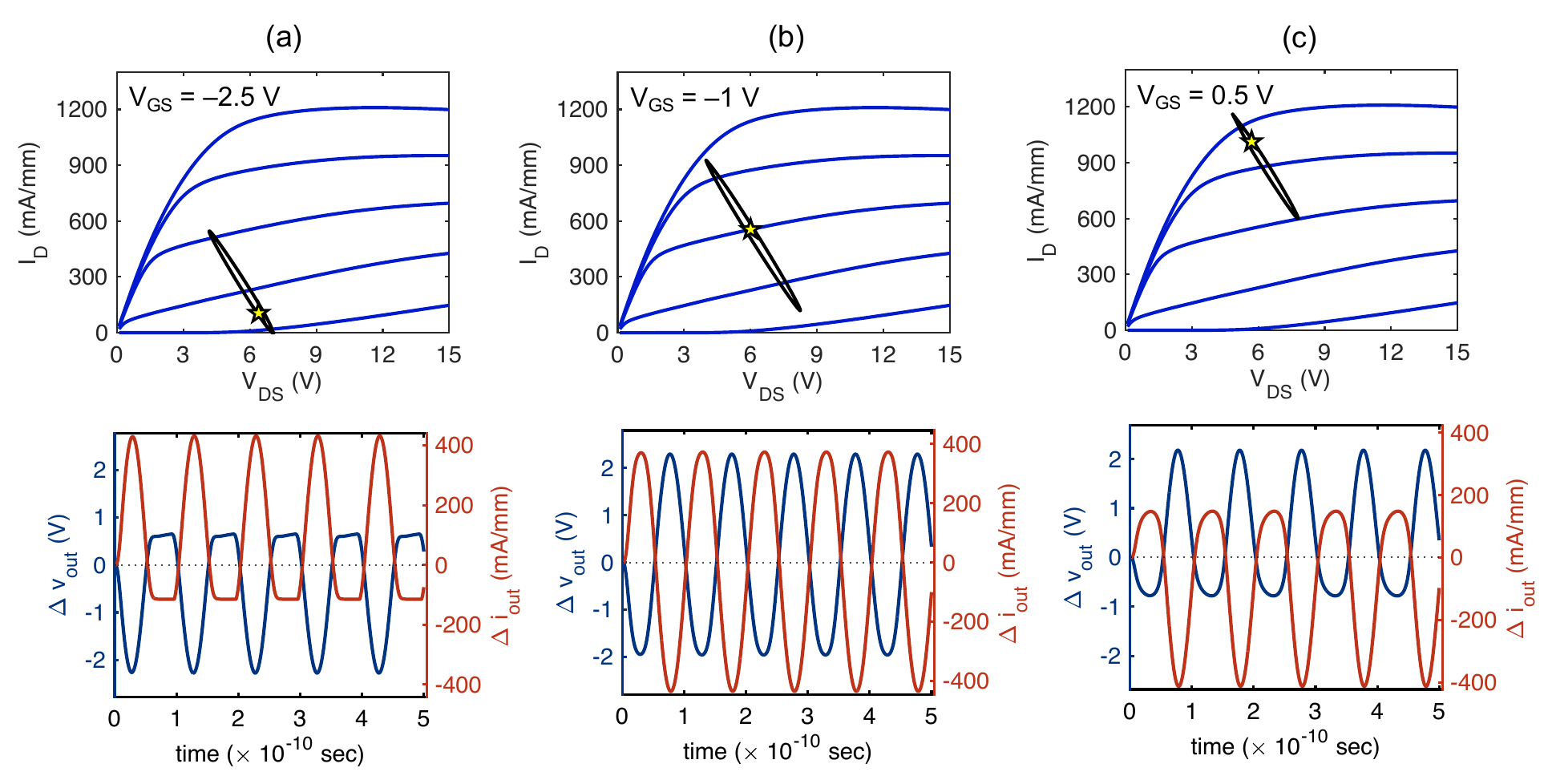}
    %\vspace{-15pt}
    \caption{Dynamic load-line (top) and $\Delta v_\textrm{out}$ and $\Delta i_\mathrm{out}$ waveforms (bottom) at (a) $V_\mathrm{GS}=-2.5$\,V, (b) $V_\mathrm{GS}=-1$\,V, and (c) $V_\mathrm{GS}=0.5$\,V for a fixed input power, $P_\mathrm{in}=7.5$\,dBm and drain bias and $V_\mathrm{DS}$ = 6\,V. At this input power, the load-line reaches the cutoff region in (a), remains between the linear and cutoff regions in (b), and extends into the linear (non-saturating) region in (c). Consequently, nonlinear behavior in the large-signal RF metrics is readily apparent in (a) and (c), but to a much lesser degree in (b).}
    \label{fig:dynamic_LL}
\end{figure*}

\vspace{-15pt}
\section{Conclusion}
\vspace{-10pt}

A systematic investigation of the role of hot LO phonons in determining the RF nonlinearity of GaN HEMTs is presented using a physics-based full-band transport framework. Large-signal simulations demonstrate that hot LO phonons significantly degrade RF performance, impacting key RF metrics such as output power, gain, power-added efficiency, OP$_\mathrm{1dB}$, and OIP3. By comparing different phonon heating scenarios, the results show that LO-phonon heating is the dominant mechanism responsible for RF compression, while acoustic phonon heating plays a comparatively smaller role. Furthermore, the analysis reveals that commonly used small-signal indicators such as the second-order transconductance derivative ($g_\textrm{m3}$) fail to reliably predict RF linearity. Devices exhibiting similar $g_\textrm{m3}$ values can still show substantial differences in large-signal RF metrics, highlighting the limitations of relying solely on transconductance derivatives for linearity assessment. Instead, the results emphasize the importance of large-signal analysis and dynamic load-line behavior in accurately capturing RF nonlinearities. Overall, this study clarifies the physical origin of RF nonlinearity in GaN HEMTs and underscores the intrinsic limitation imposed by hot LO phonons, providing important insights for the design and optimization of high-power RF GaN devices.

\vspace{-15pt}
\section{Appendix}
\vspace{-10pt}

\appendix

To understand the $P_\mathrm{{out}}-P_\mathrm{{in}}$ relationship, Fig.~\ref{fig:dynamic_LL}(a) plots the 
dynamic load line representing the trajectory of the transistor’s drain current and voltage during the RF operation, governed by the external load and the transistor's nonlinear output characteristics. 
As the input power $P_\mathrm{{in}}$ increases, the dynamic load line extends and eventually enters the non-saturation (linear or triode-like), or the cutoff region of the HEMT’s output characteristics, causing a collapse in $P_\mathrm{{out}}$ due to reduced gain and increased signal distortion. When the Q-point is roughly at the middle of the bias range ($V_\mathrm{GS}\approx -1$\,V, $V_\mathrm{DS}=6$\,V), as shown in Fig.~\ref{fig:dynamic_LL}(b), the output voltage has the maximum allowed swing without reaching the non-saturation or cutoff region, thus requiring a higher $P_\mathrm{in}$ to saturate, and resulting in a higher OP$_\mathrm{1dB}$ and OIP3. 

In the main text, the selection of the Q-point at $I_\mathrm{D} = 900\,\mathrm{mA/mm}$ and $V_\mathrm{DS} = 6\,\mathrm{V}$ is motivated by its suitability for probing the impact of LO phonons on RF performance. At this bias condition, the RF voltage swing drives the device into the knee region prior to cutoff, resembling high-bias or class-A-like operation. In contrast, biasing at lower drain currents reduces LO-phonon heating, thereby diminishing its influence on large-signal characteristics. The chosen operating regime therefore accentuates LO-phonon-induced compression, enabling a more definitive assessment of their role in RF nonlinearity.

For better linearity, the device may be engineered to have a steeper linear region, thereby preserving saturation behavior over a wider output swing. Yet, operating the device in the compression region increases the PAE of the amplifier. Further analysis of device engineering for linearity and/or efficiency is beyond the scope of this study.

\vspace{-10pt}
\begin{acknowledgments}
\vspace{-10pt}
This work was supported by AFOSR Grant No.~LRIR 24RYCOR009, DARPA Agreement No.~HR00112390072, and NSF Grant No.~ECCS-2237663.
\end{acknowledgments}

\vspace{-10pt}
\section*{Author Declarations}
\vspace{-10pt}
\subsection*{Conflict of Interest}
\vspace{-10pt}
The authors have no conflicts to disclose.
\vspace{-10pt}
\subsection*{Author Contributions}
\vspace{-10pt}
\textbf{Ankan Ghosh Dastider:} Formal analysis (lead); Investigation (equal); Validation (lead); Visualization (lead); Writing original draft (equal). \textbf{Matt Grupen:} Conceptualization (equal); Investigation (equal); Software (lead); Methodology (lead); Writing original draft (equal); Resources (equal). \textbf{Nicholas C. Miller:} Software (supporting); Writing -- review \& editing (supporting); \textbf{Shaloo Rakheja:} Conceptualization (equal); Funding acquisition (lead); Project administration (lead); Resources (equal); Supervision (lead); Writing original draft (equal).

\vspace{-10pt}
\section*{Data Availability Statement}
\vspace{-10pt}
The data that support the findings of this study are available from the corresponding author upon reasonable request.
\vspace{-10pt}

\section*{References}
\vspace{-10pt}
\bibliography{aipsamp}% Produces the bibliography via BibTeX.

@PREAMBLE{
 "\providecommand{\noopsort}[1]{}" 
 # "\providecommand{\singleletter}[1]{#1}%" 
}

@inproceedings{tang2017simulation,
  title={{Simulation of GaN HEMT with wide-linear-range transconductance}},
  author={Tang, Chenjie and Teo, Koon Hoo and Shi, Junxia},
  booktitle={2017 International Conference on Electron Devices and Solid-State Circuits (EDSSC)},
  pages={1--2},
  year={2017},
  organization={IEEE}
}

@inproceedings{inoue2013linearity,
  title={Linearity improvement of {GaN HEMT} for {RF} power amplifiers},
  author={Inoue, Kazutaka and Yamamoto, Hiroshi and Nakata, Ken and Yamada, Fumio and Yamamoto, Takashi and Sano, Seigo},
  booktitle={2013 IEEE Compound Semiconductor Integrated Circuit Symposium (CSICS)},
  pages={1--4},
  year={2013},
  organization={IEEE}
}

@article{miller2018computational,
  title={Computational study of Fermi kinetics transport applied to large-signal {RF} device simulations},
  author={Miller, Nicholas C and Grupen, Matt and Beckwith, Kris and Smithe, David and Albrecht, John D},
  journal={Journal of Computational Electronics},
  volume={17},
  number={4},
  pages={1658--1675},
  year={2018},
  publisher={Springer}
}

@inproceedings{miller2023recent,
  title={Recent advances in {GaN HEMT} modeling using fermi kinetics transport},
  author={Miller, Nicholas C and Grupen, Matt and Albrecht, John D},
  booktitle={2023 Device Research Conference (DRC)},
  pages={1--1},
  year={2023},
  organization={IEEE}
}

@article{alim2021experimental,
  title={Experimental insight into the third-order intercepts and nonlinear distortion of {GaN HEMTs}},
  author={Alim, Mohammad A and Ali, Mayahsa M and Gaquiere, Christophe},
  journal={International Journal of RF and Microwave Computer-Aided Engineering},
  volume={31},
  number={2},
  pages={e22513},
  year={2021},
  publisher={Wiley Online Library}
}

@article{shrestha2020high,
  title={High linearity and high gain performance of {N-polar GaN MIS-HEMT at 30 GHz}},
  author={Shrestha, Pawana and Guidry, Matthew and Romanczyk, Brian and Hatui, Nirupam and Wurm, Christian and Krishna, Athith and Pasayat, Shubhra S and Karnaty, Rohit R and Keller, Stacia and Buckwalter, James F and others},
  journal={IEEE Electron Device Letters},
  volume={41},
  number={5},
  pages={681--684},
  year={2020},
  publisher={IEEE}
}

@inproceedings{white2023large,
  title={Large-signal modeling of {GaN HEMTs using Fermi} kinetics and commercial hydrodynamics transport},
  author={White, E and Tunga, Ashwin and Miller, Nicholas C and Grupen, Matt and Albrecht, John D and Rakheja, Shaloo},
  booktitle={2023 Device Research Conference (DRC)},
  pages={1--2},
  year={2023},
  organization={IEEE}
}

@article{teoEmergingGaN2021,
  author  = {Teo, K. H. and Zhang, Y. and Chowdhury, N. and Rakheja, S. and Ma, R. and Xie, Q. and Yagyu, E. and Yamanaka, K. and Li, K. and Palacios, T.},
  title   = {Emerging {GaN} technologies for power, {RF}, digital, and quantum computing applications: Recent advances and prospects},
  journal = {Journal of Applied Physics},
  volume  = {130},
  number  = {16},
  pages   = {160902},
  year    = {2021},
  month   = oct,
  doi     = {10.1063/5.0061555}
}

@inproceedings{palmourWideBandgapRF2001,
  author    = {Palmour, J. W. and Sheppard, S. T. and Smith, R. P. and Allen, S. T. and Pribble, W. L. and Smith, T. J.},
  title     = {Wide bandgap semiconductor devices and {MMIC}s for {RF} power applications},
  booktitle = {International Electron Devices Meeting},
  pages     = {17.4.1--17.4.4},
  year      = {2001},
  doi       = {10.1109/IEDM.2001.979517}
}

@article{nagyLinearityGaNHEMT2003,
  author  = {Nagy, W. and Brown, J. and Borges, R. and Singhal, S.},
  title   = {Linearity characteristics of microwave-power {GaN HEMTs}},
  journal = {IEEE Transactions on Microwave Theory and Techniques},
  volume  = {51},
  number  = {2},
  pages   = {660--664},
  year    = {2003},
  month   = feb,
  doi     = {10.1109/TMTT.2002.807684}
}

@inproceedings{jenkinsLinearityAlGaN2001,
  author    = {Jenkins, T. and Kehias, L. and Parikh, P. and Wu, Y.-F. and Chavarkar, P. and Moore, M. and Mishra, U.},
  title     = {Linearity of high {Al}-content {AlGaN/GaN HEMTs}},
  booktitle = {Device Research Conference},
  pages     = {201--202},
  year      = {2001},
  doi       = {10.1109/DRC.2001.937931}
}

@article{liMonteCarloAlGaN2000,
  author  = {Li, T. and Joshi, R. P. and Fazi, C.},
  title   = {{Monte Carlo} evaluations of degeneracy and interface roughness effects on electron transport in {AlGaN–GaN} heterostructures},
  journal = {Journal of Applied Physics},
  volume  = {88},
  number  = {2},
  pages   = {829--837},
  year    = {2000},
  doi     = {10.1063/1.373744}
}

@article{bajajTransportGaNHEMT2015,
  author  = {Bajaj, S. and Shoron, O. F. and Park, P. S. and Krishnamoorthy, S. and Akyol, F. and Hung, T.-H. and Reza, S. and Chumbes, E. M. and Khurgin, J. and Rajan, S.},
  title   = {Density-dependent electron transport and precise modeling of {GaN} high electron mobility transistors},
  journal = {Applied Physics Letters},
  volume  = {107},
  pages   = {153503},
  year    = {2015},
  doi     = {10.1063/1.4933181}
}

@article{chenBellShapeGM2016,
  author  = {Chen, C.-H. and Sadler, R. and Wang, D. and Hou, D. and Yang, Y. and Yau, W. and Sutton, W. and Shim, J. and Wang, S. and Duong, A.},
  title   = {The causes of {GaN HEMT} bell-shaped transconductance degradation},
  journal = {Solid-State Electronics},
  volume  = {126},
  pages   = {115--124},
  year    = {2016},
  doi     = {10.1016/j.sse.2016.09.005}
}

@article{palaciosDynamicAccessResistance2005,
  author  = {Palacios, T. and Rajan, S. and Chakraborty, A. and Heikman, S. and Keller, S. and DenBaars, S. P. and Mishra, U. K.},
  title   = {Influence of the dynamic access resistance in the $g_\mathrm{m}$ and $f_\mathrm{T}$ linearity of {AlGaN/GaN HEMTs}},
  journal = {IEEE Transactions on Electron Devices},
  volume  = {52},
  pages   = {2117--2123},
  year    = {2005},
  doi     = {10.1109/TED.2005.856180}
}

@article{ridleyHotPhononVelocity2004,
  author  = {Ridley, B. K. and Schaff, W. J. and Eastman, L. F.},
  title   = {Hot-phonon-induced velocity saturation in {GaN}},
  journal = {Journal of Applied Physics},
  volume  = {96},
  pages   = {1499--1502},
  year    = {2004},
  doi     = {10.1063/1.1762999}
}

@article{grupenHeatFlowTransport2009,
  author  = {Grupen, M.},
  title   = {An alternative treatment of heat flow for charge transport in semiconductor devices},
  journal = {Journal of Applied Physics},
  volume  = {106},
  pages   = {123702},
  year    = {2009},
  doi     = {10.1063/1.3270404}
}

@article{juangTransportAlGaNGaN2003,
  author  = {Juang, J. R. and Huang, T.-Y. and Chen, T.-M. and Lin, M.-G. and Kim, G.-H. and Lee, Y. and Liang, C.-T. and Hang, D. R. and Chen, Y. F. and Chyi, J.-I.},
  title   = {Transport in a gated $\mathrm{Al}_{0.18}\mathrm{Ga}_{0.82}\mathrm{N}$/{GaN} electron system},
  journal = {Journal of Applied Physics},
  volume  = {94},
  pages   = {3181--3184},
  year    = {2003},
  doi     = {10.1063/1.1594818}
}

@article{wuTransconductanceCollapse2005,
  author  = {Wu, Y.-R. and Singh, M. and Singh, J.},
  title   = {Sources of transconductance collapse in {III-V} nitrides},
  journal = {IEEE Transactions on Electron Devices},
  volume  = {52},
  pages   = {1048--1054},
  year    = {2005},
  doi     = {10.1109/TED.2005.848084}
}

@article{russoSourceGateDistance2007,
  author  = {Russo, S. and Di Carlo, A.},
  title   = {Influence of the source–gate distance on {AlGaN/GaN HEMT} performance},
  journal = {IEEE Transactions on Electron Devices},
  volume  = {54},
  pages   = {1071--1075},
  year    = {2007},
  doi     = {10.1109/TED.2007.894614}
}

@inproceedings{joglekarVTLinearity2017,
  author    = {Joglekar, S. and Radhakrishna, U. and Piedra, D. and Antoniadis, D. and Palacios, T.},
  title     = {Large signal linearity enhancement of {AlGaN/GaN HEMTs} by device-level $V_T$ engineering},
  booktitle = {IEEE International Electron Devices Meeting},
  year      = {2017},
  doi       = {10.1109/IEDM.2017.8268457}
}

@article{azadMultimetalGateGaN2023,
  author  = {Azad, Md. T. and Hossain, T. and Sikder, B. and Xie, Q. and Yuan, M. and Yagyu, E. and Teo, K. H. and Palacios, T. and Chowdhury, N.},
  title   = {{AlGaN/GaN}-Based Multimetal Gated HEMT With Improved Linearity},
  journal = {IEEE Transactions on Electron Devices},
  volume  = {70},
  pages   = {5570--5576},
  year    = {2023},
  doi     = {10.1109/TED.2023.3311422}
}

@article{tarakjiLargeSignalIII-N2003,
  author  = {Tarakji, A. and Fatima, H. and Hu, X. and Zhang, J.-P. and Simin, G. and Khan, M. A. and Shur, M. S. and Gaska, R.},
  title   = {Large-signal linearity in {III-N MOSDHFETs}},
  journal = {IEEE Electron Device Letters},
  volume  = {24},
  pages   = {369--371},
  year    = {2003},
  doi     = {10.1109/LED.2003.813355}
}

@article{palaciosEmodeGaN2006,
  author  = {Palacios, T. and Suh, C.-S. and Chakraborty, A. and Keller, S. and DenBaars, S. P. and Mishra, U. K.},
  title   = {High-performance {E-mode AlGaN/GaN HEMTs}},
  journal = {IEEE Electron Device Letters},
  volume  = {27},
  pages   = {428--430},
  year    = {2006},
  doi     = {10.1109/LED.2006.874761}
}

@article{marinoDislocationsGaNHEMT2010,
  author  = {Marino, F. A. and Faralli, N. and Palacios, T. and Ferry, D. K. and Goodnick, S. M. and Saraniti, M.},
  title   = {Effects of threading dislocations on {AlGaN/GaN HEMTs}},
  journal = {IEEE Transactions on Electron Devices},
  volume  = {57},
  pages   = {353--360},
  year    = {2010},
  doi     = {10.1109/TED.2009.2035024}
}

@article{matulionisPlasmonHeat2009,
  author  = {Matulionis, A. and Liberis, J. and Matulionienė, I. and Ramonas, M. and Šermukšnis, E. and Leach, J. H. and Wu, M. and Ni, X. and Li, X. and Morkoç, H.},
  title   = {Plasmon-enhanced heat dissipation in {GaN}-based two-dimensional channels},
  journal = {Applied Physics Letters},
  volume  = {95},
  pages   = {192102},
  year    = {2009},
  doi     = {10.1063/1.3261748}
}

@article{liberisHotPhononLifetime2014,
  author  = {Liberis, J. and Ramonas, M. and Šermukšnis, E. and Sakalas, P. and Szabo, N. and Schuster, M. and Wachowiak, A. and Matulionis, A.},
  title   = {Hot-phonon lifetime in {AlGaN/GaN} channels},
  journal = {Semiconductor Science and Technology},
  volume  = {29},
  pages   = {045018},
  year    = {2014},
  doi     = {10.1088/0268-1242/29/4/045018}
}

@article{matulionisHotPhononTemperature2003,
  author  = {Matulionis, A. and Liberis, J. and Matulionienė, I. and Ramonas, M. and Eastman, L. F. and Shealy, J. R. and Tilak, V. and Vertiatchikh, A.},
  title   = {Hot-phonon temperature and lifetime in biased {AlGaN/GaN} channels},
  journal = {Physical Review B},
  volume  = {68},
  pages   = {035338},
  year    = {2003},
  doi     = {10.1103/PhysRevB.68.035338}
}

@article{dastiderFullBandGaN2026,
  author  = {Dastider, A. G. and Grupen, M. and Tunga, A. and Rakheja, S.},
  title   = {Physics-based full-band {GaN} HEMT simulation suggests upper bound of {LO} phonon lifetime},
  journal = {Journal of Applied Physics},
  volume  = {139},
  pages   = {074502},
  year    = {2026},
  doi     = {10.1063/5.0315424}
}

@article{grupenEnergyTransportGaAs2011,
  author  = {Grupen, M.},
  title   = {Energy transport model with full band structure for {GaAs} electronic devices},
  journal = {Journal of Computational Electronics},
  volume  = {10},
  pages   = {271--290},
  year    = {2011},
  doi     = {10.1007/s10825-011-0364-9}
}

@ARTICLE{grupenFullWaveGaN2016,
  author={Grupen, M.},
  title={{GaN} High Electron Mobility Transistor Simulations With Full Wave and Hot Electron Effects}, 
  journal={IEEE Transactions on Electron Devices},
  volume={63},
  number={8},
  pages={3096-3102},
  year={2016},
  doi={10.1109/TED.2016.2581591}
}

@article{tungaTCADComparison2022,
  author  = {Tunga, A. and Li, K. and White, E. and Miller, N. C. and Grupen, M. and Albrecht, J. D. and Rakheja, S.},
  title   = {Comparison of hydrodynamic {TCAD} solver and Fermi kinetics transport for {GaN HEMTs}},
  journal = {Journal of Applied Physics},
  volume  = {132},
  pages   = {225702},
  year    = {2022},
  doi     = {10.1063/5.0118104}
}

@article{barmanNonequilibriumPhonons2004,
  author  = {Barman, S. and Srivastava, G. P.},
  title   = {Long-wavelength nonequilibrium optical phonon dynamics in cubic and hexagonal semiconductors},
  journal = {Physical Review B},
  volume  = {69},
  pages   = {235208},
  year    = {2004},
  doi     = {10.1103/PhysRevB.69.235208}
}

@article{srivastavaHotPhononOrigin2008,
  author  = {Srivastava, G. P.},
  title   = {Origin of the hot phonon effect in group-{III} nitrides},
  journal = {Physical Review B},
  volume  = {77},
  pages   = {155205},
  year    = {2008}
}

@book{ridleyQuantumProcesses2013,
  author    = {Ridley, B. K.},
  title     = {Quantum Processes in Semiconductors},
  publisher = {Oxford University Press},
  year      = {2013}
}

\end{document}